# Amidst Uncertainty -- or Not?
# Decision-Making in Early-Stage Software Startups

Kai-Kristian Kemell[1], Eveliina Ventilä[1], Petri Kettunen[2], and Tommi Mikkonen[2]

[1] University of Jyväskylä, Jyväskylä, Finland.
[2] University of Helsinki, Helsinki, Finland.

**Abstract.** Startups are an important part of society's development. New, innovative startups challenge traditional companies to innovate, create templates for flexible and risk-taking business, and innovate the sector's business structures. It is commonly claimed that the initial stages of any start-up business are dominated by continuous, extended uncertainty, in an environment that has even been described as chaotic. Consequently, decisions are made in uncertain circumstances, so making the right decision is crucial to successful business. However, little currently exists in the way of empirical studies into this supposed uncertainty, the assumption of which we now challenge based on empirical data. In this paper, we shed light on decision-making in early-stage software startups by means of a single, in-depth case study. Based on our data, we argue that software startups in fact do not work in a chaotic environment, nor are most of the decisions made in software startups characterized by uncertainty.

## 1 Introduction

Startups are considered an important part of society's development. New, innovative startups challenge traditional companies to innovate, create templates for flexible and risk-taking business, and innovate the sector's business structures [15]. In general, startup culture is seen as being driven by innovation, agility and growth. On the other hand, startups are considered to operate with scarce resources and in uncertain conditions where failure is likelier than success. It is commonly claimed that the initial stages of any startup business are dominated by continuous uncertainty [22].

In software-intensive startups, or simply software startups, the role of software in the final offering may vary from being the core product to merely serving as an enabler or support of the main business idea [23]. For example, Uber is considered a software startup, as it delivers its value through a software application. In integrated products, the software may be embedded in the hardware parts of the products. Consequently, the sources of uncertainty may potentially be many-sided and complex.

Various other characteristics have been attributed to startups in an attempt to differentiate them from other business organizations. These include the aforementioned lack of resources, as well as such attributes as (1) highly reactive, (2) innovation, (3) uncertainty, (4) rapidly evolving, (5) time-pressure, (6) third party dependency, (7) small team, (9) one product), (10) low-experienced team, (11) new company, (12) flat organization, (13) highly risky, (14) not self-sustained, and (15) little working history [15]. While these characteristics are seldom directly used to define what a software startup is, they are often listed in various contexts while discussing software startups are, in the absence of a widely accepted definition for a software startup.

Indeed, the search for a comprehensive definition of a startup continues. Recently, the empirical evidence behind many of the aforementioned characteristics has been questioned by Klotins [9] who argue that the original sources behind these claims are hardly conclusive. Through a literature review, they contest the uniqueness of software startups in relation to the following attributes: riskiness or failure rates, lack of software engineering experience, innovativeness, market-related time pressure, and lack of resources. This has sparked further discussion into what software startups actually are and how they truly differ from other business organizations.

Decision-making is another area where little is known empirically about software startups. Unterkalmsteiner et al. [22] have recommended further research into the topic. The current understanding is that software startups work amidst uncertainty that has even been described as a chaotic environment [3, 14, 15].

To provide empirical evidence into this on-going debate on software startup characteristics, we study software startup decision-making in relation to the uncertainty attributed to software startup in this paper. We challenge this established view, to an extent, and claim that most of the decisions a software startup entrepreneur has to make can be made analytically and in a logical fashion, and that startups are not notably unique in terms of the uncertainty they do experience. We thus study decision-making in software startups by means of a single, in-depth case study executed in an ethnography-inspired fashion. We utilize the Cynefin framework [12] for decision-making to study decision-making in the case startup and the uncertainty associated with it. Specifically, we seek to tackle the following research question in doing so:

**RQ:** How can we characterize the context software startups operate in, in relation to decision-making, and more specifically uncertainty in decision-making?

The rest of this paper is structured as follows. In Section 2, we discuss the background and related work, and in Section 3 we introduce the framework used in the analysis of our data. In Section 4, we introduce our case study. In Section 5, we present our results, which we then discuss in Section 6. Section 7 concludes the paper.

## 2      Background

Little currently exists in the way of studies focusing on decision-making specifically in the context of software startups [22]. On the other hand, organizational decision-making is a well-established area of research spanning various disciplines, including psychology, various economic disciplines, and software engineering. Areas of research related to it are similarly numerous. E.g., business intelligence focuses on using data to support organizational decision-making.

Research on decision-making in the area of New Product Development (NPD) can be considered to be related to the context of software startups, at least on a conceptual level. Software startups are seen as searching for new business models [22], whereas conventional business organizations, in this line of reasoning, *execute* business models. E.g. a new restaurant is not a startup because its general business model is very well established. Software startups also make various decisions regarding software development in particular, such as deciding on development practices [15].

In a literature review spanning multiple disciplines, Krisnan and Ulrich [11] presented a list of decisions related to setting up a product development project, which is interesting from the point of view of this study. In their literature review, they [11] list various higher-level decisions related to product development that an organization needs to make when starting a product development project. Many of these decisions such as "which technologies will be employed in the product(s)?" can be considered potentially relevant from the point of view of software startups as well, while some decisions are far more relevant to larger, more established companies, e.g. "will a functional, project, or development matrix organization be used?".

The extant literature in this area has focused on conventional business organizations rather than (software) startups. The argument used to justify studies into software startups is that they differ from said conventional businesses, making the findings of such studies not (fully) applicable to them. Thus, in debating whether they really do notably differ from other business organizations, empirical studies into the defining characteristics of software startups are needed.

## 3      Research Framework: Cynefin

Adhering to the fundamentals of qualitative research in the context of case studies (as e.g. discussed by Järvinen ([8], p. 75), we utilize existing literature, specifically the Cynefin framework, to analyze our data. The Cynefin framework (Fig 1) is a decision-making tool from the field of knowledge management and complexity science [12, 19, 20]. It a sense-making tool intended to help its users make sense of the current context around them while making decisions. It presents a typology for decisions and provides guidelines for making decisions of each type. It is based on three assumptions: (1) the assumption of order; (2) the assumption of rational choice, and (3) the assumption of internal capability [12].

More specifically, the assumption of order refers to the idea that there are underlying relationships between causes and effects in a given context. Rational choice, in this context, means that human actors will make a rational choice in any situation based on what effect they perceive their action will have. Finally, internal capability is related the assumption that external actions are a result of intentional behavior.

The Cynefin framework splits decisions into five domains. These domains are based on the aforementioned assumption of order, i.e. perceived causality of cause and effect. Each domain contains characteristics describing decisions of that domain, as well as recommended actions for decision-making.

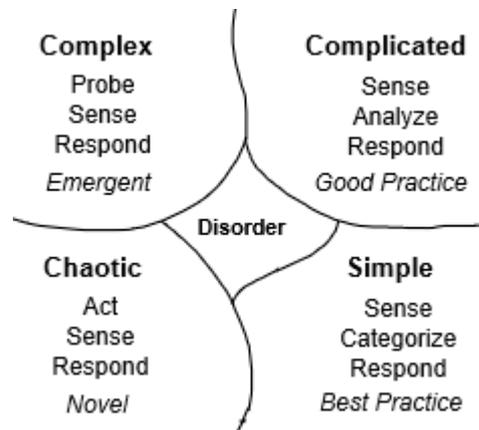

**Fig. 1.** The Cynefin Framework [12]

In the *simple* domain, cause and effect are clearly and easily discernible. It is a well-understood domain that has established best practices. The actions associated with the simple domain are sense, categorize, and respond. I.e. one should establish the facts (sense), categorize the context, and then respond with a best practice. For example, a bike chain dislodging while cycling results in what is a simple decision with a clear cause and effect for most people: manually adjust it back into its place.

In the *complicated* domain, cause and effect are still clear and discernible, but only as a result of (expert) analysis. The actions associated with the complicated domain are sense, analyze, and respond. Similar to the simple domain, one should first establish the facts (sense), before conducting an expert analysis of the situation, and finally respond with a good practice. Complicated situations are characterized by the need for expert analysis, whereas simple decisions are clerical decisions requiring little expertise. For example, if your car starts making clanking sounds, you might assume something is wrong, but peeking under the hood of the car will tell you very little unless you are an expert in the domain. Thus, an analysis conducted by an expert is needed.

On the next level of un-orderedness lies the *complex* domain. In this domain, causalities are not apparent until cause and effect can be analyzed in retrospect. This poses challenges for decision-making, necessitating experimentation in order to decide on the (seemingly) best course of action. Indeed, in this domain, the recommended actions are probe, sense, and respond. One would start by experimenting (probing) in order to establish some facts (sense), before acting in a suitable way in response.

Finally, the *chaotic* domain is characterized by the lack of perceivable cause and effect relations. This domain is characterized by time pressure, which leaves no room for probing or analysis and forces one to act ad hoc. The recommended actions are act, sense, and respond. I.e. the time pressure forces one to act quickly, and the results of that action can then be evaluated in retrospect in an attempt to establish facts (sense), after which one has to make a decision again (respond). This process continues iteratively until it is possible to make enough sense of the situation to exit the chaotic domain. Various crisis management situations are examples of chaos, as the effects of typical risk management practices tend to vary greatly based on context.

In addition to the aforementioned four domains, *disorder* refers to a situation where the domain is unclear. Disorder necessitates an analysis of the situation in order to understand what domain the problem is. To this end, it may not be clear to the individual when they are in the disorder domain.

The Cynefin framework has been utilized in various fields. It has been used to study decision-making in management science (e.g. [4, 5]). In the context of IT disciplines, it has seen some utilization in information systems [7], as well as in SE in relation to software startups [14]. Moreover, Paternoster et al. [15] noted that the framework could be useful in explaining "the orientation of software startups towards flexible and creative development approaches". They also note that they consider software startups to operate between the complex and chaotic domain in Cynefin, which is something we challenge in this study

## 4   Research Methodology and Case Description

To study decision-making in software startups, we conducted a single in-depth case study of a Finnish software startup. The study was conducted in an ethnography-inspired fashion, with one of the authors working as the founder of a software startup for six months while collecting data. Despite this being the research setting, the case startup was a real-life startup not founded for the purpose of carrying out this study.

The software startup studied in this paper was founded in early 2018 by an inexperienced founder. At the time of the start of the data collection, the software startup in question was a very early-stage startup with only the founder on the team. The software startup had an idea in the form of a problem it was going to tackle, as well as a solution it was going to produce in order to tackle the problem. Past that, the startup did not yet have a name, a team, a website, or a single line of code written.

The startup, from here-on-out Startup A, produces a software service, delivered via a dedicated hardware solution, for supporting social interaction in public events. More specifically, a wristband that would act as a replacement for business cards. At the time of the founding of the company, it was unclear whether the company would develop only the software or (parts of) the hardware as well. As the founder had no technical SE skills, this would be decided later.

Data from the case was collected for six months, between 30 April 2018 and 30 October 2018, in the form of video diary entries. Each evening, the founder was to produce a video recording detailing each decision they had made that day in relation to Startup A. In each entry, they listed the decisions made since the previous entry, along with detailing the current situation the startup was in at the time.

In practice, the entries were occasionally produced every other day if no decisions were made on a particular day. Furthermore, the founder put the project on hold for the summer between June 27th and August 5th, and at some points during the data collection, avoided working in the weekends. After the conclusion of the data collection, the video diary entries were transcribed. The transcripts included a full word-to-word transcript of each entry and a list of every decision discussed in each entry.

From the transcripts, one of the authors extracted all the decisions discussed in the video diary entries, and this list of decisions was then analyzed through the Cynefin framework (section 3), using the full set of data to understand the context Startup A was in. A total of 136 decisions were analyzed. Each decision was evaluated and placed into the corresponding domain according to the criteria described in Table 1.

In order to increase the rigor of the analysis, we followed the protocol below:
1. Author A (case startup founder) categorizes the data according to the Cynefin model and provides reasoning for their categorizations in one or two sentences per decision.
2. Author B categorizes the data independently, without studying the results of Author A's analysis.
3. Author B compares the results of their analysis with that of Author A.
4. Decisions classified into the same category by Authors A and B are included for analysis (89 decisions, 65,4% of the total 136).
5. Author B studies the reasoning provided by Author A in the case of conflicting classifications (47 decisions, 34,6% of the total 136), and either changes their classifications by agreeing with Author A or continues to disagree.
6. Author C discusses the remaining conflicts with Author B (22 conflicts, 16,2% of the total 136 decisions).
7. The remaining conflicts are classified based on the consensus reached by Authors B and C

For the purpose of the analysis, the criteria depicted in Table 1 (below) were used to assign the decisions into domains in the Cynefin framework. If a decision did not clearly belong into a specific domain, the criteria were used to sort it based on which domain's criteria it fulfilled the most.

**Table 1.** Criteria for Assigning Decisions into the Cynefin Main Domains.

|  | Decision speed | Effects observable | Potential decisions | Key action |
|---|---|---|---|---|
| **Simple** | Fast | Immediately or quickly | Usually one correct one | Categorize |
| **Complicated** | Slow | Quickly or slight delay | Multiple potentially correct ones | Analyze |
| **Complex** | (Very) Slow | Slowly | Numerous, difficult to choose good ones | Experiment |
| **Chaotic** | Fast | Immediately or quickly | No correct option, minimize risks | Act |

## 5 Results

This section is split into five subsections according to the domains of the Cynefin framework used for data analysis (simple, complicated, complex, chaotic, and disorder). In our analysis, we highlight key observations in the form of Primary Empirical Conclusions (PEC), which we then discuss further in the discussion section. Each subsection contains five examples of the decisions of that domain, while the full list of decisions can be found on FigShare[1].

### 5.1 Simple Domain

Most of the decisions recorded by the founder of Startup A during the data collection period were simple in nature, according to the Cynefin framework. In total, 63 out of 136 decisions were simple. Indeed, an early-stage startup faces many tasks that can be considered universal for early-stage startups, or simply beneficial activities even if not mandatory. These simple decisions (examples in Table 2) included setting up a company website and social media profiles, creating a logo, taking photos of the team for the web content, and coming up with a company name. This underlines the large number of menial activities any early-stage software startup has to carry out.

Past the decisions that could be considered relevant for nearly any software startup, Startup A was also faced with many context-specific but nonetheless simple decisions. These included setting up meetings with organizations belonging to the local startup ecosystem, dressing up for said meetings, and creating a calendar for the team in order to keep track of events.

**Table 2.** Examples of Simple Decisions.

| **Simple Decisions** |
|---|
| Accepting an office space that was offered to the startup (for free) |
| Setting up social media accounts for the startup |
| Focusing on acquiring funding |
| Creating a common team calendar |
| Ordering a rollup |

However, many of these simple decisions led to further decisions that were complicated or even complex in nature. The decision to set up a company is simple because it is a best practice, but decisions such as what type of content to put there require analysis. This was also the case in relation to funding: Startup A obviously needed funding, but further decisions on that front quickly turned complex.

### 5.2 Complicated Domain

The second most prominent domain was the complicated one: 46 out of 136 decisions made by Startup A were considered to have been complicated (examples in Table 3). The decisions in the complicated domain required expertise from various areas of business and IT, underlining the multidisciplinary nature of software startups. Complicated decisions need expert analysis, which can prove challenging especially for an early-stage software startup working with a small team with limited expertise.

While Startup A was able to successfully execute many of the complicated decisions, these decisions were nonetheless taxing resource-wise. The inexperienced team of the startup did not know how to find a limited company or which forms of funding were available to them in the local startup ecosystem. Many of these decisions would have been less taxing to execute for a more experienced team with prior entrepreneurship experience, lending more resources towards progressing the business.

---
[1] https://doi.org/10.6084/m9.figshare.8298008.v1.

**PEC1:** Inexperience increases the workload of a software startup team, contributing to the lack of resources typically associated with software startups

Furthermore, many of the complicated decisions were in fact related to resource allocation. For example, the team regularly weighed between different funding options and made decisions regarding funding applications, trying to analyze which options were the least likely to end up being a waste of resources. They considered their lack of financial resources to be in various ways detrimental. It not only potentially undermined team morale, but also forced them to continuously devote resources towards securing funding. Many of the funding sources required formal funding applications which took notable amounts of time to prepare. Alternatively, they had to keep pitching their idea to potential investors, which also required preparation.

**PEC2:** A lack of financial resources contributes towards a lack of resources in terms of person-hours. I.e., a software startup in a need of funding must devote notable amounts of person-hours towards acquiring funding

Startup A was able to tackle some of the complicated issues using the resources of their own team. Many of the complicated decisions, however, ultimately ended up requiring expertise the team did not possess. In some cases (e.g. designing a logo) the team decided to study the skills required to carry out the task themselves. In other cases, such as engineering the service (programming), the team decided to enlist a new member as opposed to e.g. having the founder study programming, which was deemed unreasonably resource intensive. Issues related to the expertise of the team of Startup A were prominent in their decision-making, although many of the more profound issues in relation to the team's expertise were complex rather than complicated.

**Table 3.** Examples of Complicated Decisions.

| Complicated Decisions |
|---|
| Acquiring a domain (which, and how) |
| Practicing public speaking |
| Deciding on different pricing models |
| Applying even for those forms of funding that are difficult to obtain |
| Studying different funding options |

### 5.3 Complex Domain

The complex domain relates most to the uncertainty associated with software startups. Only 21 decisions out of 136 (15,4%) could be considered as complex in the data collected, listed in Table 4. Complex decisions are decisions that require experimentation, and which cannot be consistently solved with existing good or best practices. Thus, the remaining 115 decisions (84,6%) were decisions that *could* be solved through existing good or best practices.

**PEC3:** Only a small portion of the decisions software startups make requires experimentation as opposed to analytical decision-making based on existing good or best practices

Many of the complex decisions (examples in Table 4) in Startup A were related to the lack of financial resources they experienced. Startup A was constantly balancing between different funding options with no clear way to determine which one was the most likely one to result in receiving funding. This resulted in a situation where they were forced to pursue the funding alternatives that would have yielded funding the fastest, even if they did not consider them to be desirable funding options.

Startup A also had to continuously devote resources towards finding new prospective sources of funding after exhausting different funding options, e.g. those recommended by their mentors. In search of funding, they participated in various events, which took considerable amounts of resources. Some of the funding sources also had conflicting requirements: one source of funding required a limited company to have been founded, while another one specifically required that a limited company did *not* exist around the idea yet, adding further complexity to the issue.

**PEC4:** A lack of financial resources notably increases the level of uncertainty experienced by a software startup

Aside from funding issues, Startup A operated in a complex environment in relation to their service. They struggled to devise a software or hardware MVP to test their idea with out on the field. They discovered that they lacked the expertise to develop the hardware themselves and took to looking into existing hardware solutions that they could develop software for. While struggling to develop the service, they ideated alternative ways to validate their idea through other types of demos or user surveys. In general, finding the suitable team in terms of size, expertise, and person fit, was a long-standing and complex problem for Startup A, which also reflected on their decisions related to the team in the preceding two domains.

**PEC5:** A lack of technical know-how in the team is a critical issue for software startups and increases the uncertainty experienced by a software startup

The remaining complex decisions made by Startup A largely stemmed from their lack of funding. These involved decisions relating to how long the team could be expected to keep going without funding, and at which point should they consider quitting in case they could not secure outside funding. The team was not concerned about funding in terms of salaries, being passionate about their idea, but felt that they could not take the development of the service further without acquiring funding for hardware. They were concerned about the viability of their idea but struggled to find ways to validate it without a functional prototype.

**Table 4.** Examples of Complex Decisions.

| Complex Decisions |
|---|
| How long should the project be continued (without funding) |
| Validating the most important features (of the future system) |
| Deciding on funding source based on where to get funding the fastest |
| Carrying on with the endeavor despite a similar service already existing in Europe (i.e., a competitor might expand to the local region) |
| Gathering a team (what capabilities are needed, team size, etc.) |

### 5.4 Chaotic Domain

No decisions were categorized as chaotic. Occasionally, the situation Startup A was in in terms of funding was close to what could be considered a chaotic situation for decision-making. However, at no point did Startup A experience clear time pressure that necessitated acting without experimenting or analyzing the situation.

As the team was not being paid salaries, and had no history of having had funding, the team was used to the situation. The failures of Startup A to obtain funding did not therefore result in any dramatic crisis at any point, as the lack of funding had been the norm for them. It is, nonetheless, arguably possible for a software startup to find itself in a chaotic situation in relation to funding e.g. upon failing to pay salaries.

**PEC6:** Software startups do not operate in a predominantly chaotic environment

### 5.5 Disorder

Disorder is highly context-specific and applies when there is no clear understanding of the current context. It is difficult to categorize disorder retrospectively from decisions that have already been made. However, we consider general-level personal time management related decisions to largely be related to disorder. The founder discussed a number of low detail personal time management decisions that we considered to be prone to disorder: not working the weekends, prioritizing the startup, and conversely, prioritizing university studies and occasionally prioritizing personal life.

These decisions are high-level decisions that consist of various sub-decisions that are then made on a case-by-case basis. One may wish to prioritize work over personal life, but in the case of a sudden, unforeseen event such as the start of a new relationship, the situation may change from simple to complex very quickly. To this end, one may not be aware of which domain one is in at a given time in relation to personal time management, depending, e.g., on their personal time management skills.

## 6    Discussion

We have underlined our Primary Empirical Conclusions (PEC) in Table 5. In this section, we discuss each of them in relation to extant literature.

Starting with the alleged uncertainty of software startups, one of the papers that describes software startups as operating in a chaotic environment dates back to 1998 [3]. Back in the late 1990s and early 2000s, startup entrepeneurship as a phenomenon was still emerging and little guidance in terms of methods, practices, or mentorship existed. Today, software startups are numerous and various past startups have gone on to grow into mature companies. We now have data from both practice and research that has furthered our understanding of software startups. This body of knowledge includes methods such as the lean startup [17] as well as various singular practices (e.g. the five whys). Similarly, startup entrepreneurship is taught in universities.

Table 5. Primary Empirical Conclusions Based on Analysis of the Data

| # | PEC description (from analysis section) |
|---|---|
| 1 | Inexperience increases the workload of a software startup team, contributing to the lack of resources typically associated with software startups |
| 2 | A lack of financial resources contributes towards a lack of resources in terms of person-hours. I.e., a software startup in a need of funding must devote notable amounts of person-hours towards acquiring funding |
| 3 | Only a small portion of the decisions software startups make requires experimentation as opposed to analytical decision-making based on existing good or best practices |
| 4 | A lack of financial resources notably increases the level of uncertainty experienced by a software startup |
| 5 | A lack of technical know-how in the team is a critical issue for software startups and increases the uncertainty experienced by a software startup |
| 6 | Software startups do not operate in a predominantly chaotic environment |

This development has resulted in a situation where software startups have various resources that can help them operate more systematically, reducing the level of uncertainty. Indeed, in our data, we identified no chaotic decisions to have been made by the case startup (PEC6). This finding contradicts Nguyen-duc et al. [14] and Paternoster et al. [15] who suggested that software startups do largely operate in a chaotic environment, specifically in the context of the Cynefin framework.

Arguably though, a software startup with a history of funding or one that has operated under considerable amounts of personal funding may experience chaos if its funding runs out. If such a startup has been paying its team members salaries, a sudden lack of financial resources may drive the startup into a chaotic situation. Chaos is perhaps also possible in relation to the SE aspect of a startup in a software startup that has managed to progress further with the development of its service.

Moreover, software startups do not seem to operate under as much uncertainty as is typically attributed to them. In our analysis, we discovered that only 15,4% of the decisions the case startup made could be considered to have been complex, i.e. requiring experimentation. In other words, the majority of the decisions (84,6%) Startup A made were analytical in nature and could be solved by utilizing existing good and best practices, leaving little room for objective uncertainty (PEC3). Startup A was also faced with most of the decisions discussed by Krisnan and Ulrich [11] in relation to starting a new product development project. This provides further support for the idea that software startups are not as unique in relation to decision-making as has been conventionally thought, although no formal comparison with this particular study or any other past study on organizational decision-making was made in the analysis.

The uncertainty experienced by the case startup stemmed from two main sources: lack of funding and lack of technical know-how in the team. In relation to technical know-how, extant research has underlined capabilities required in early-stage software startup teams, highlighting the importance of hardware capabilities where applicable, as well as the importance of having those capabilities present from the start [18]. Throughout the data collection period, Startup A struggled to find the various capabilities required to carry out their idea. However, problems related to securing technical know-how are not unique to startups, as e.g. underlined by the current lack of programmers in the Finnish IT field in general[2].

The failure of Startup A to secure hardware capabilities from the beginning began to quickly cause issues. Unable to develop their own hardware, Startup A was forced to experiment with existing hardware products and struggled to create a functional prototype of the service (PEC5). This also furthered their lack of resources, as they required funding to better experiment with various hardware products. Ultimately, this resulted in an unreasonably long development cycle for an initial version of the product, which Klotins et al. [10] consider a key anti-pattern in software startups.

Lack of resources in relation to software startups, on the other hand, has recently been a topic of discussion in the area. Whereas past research has considered software startups to be unique in their lack of resources, Klotins [9] recently argued that the literature in fact does not provide convincing arguments to support the uniqueness of software startups in this regard. Our findings, on the other hand, support the idea that especially early-stage software startups experience the lack of resources (financial, person-hours, physical) differently compared to established organizations.

Startup A operated in a complex environment in relation to funding. They devoted notable amounts of time resources towards finding new prospective sources of funding and then pursuing them. Moreover, due to the small team size, the time resources spent on searching for funding were time resources not spent on e.g. developing the service. An established organization may struggle with financial resources, but their ways of tackling the problem will be vastly different (e.g. aiming to improve sales numbers as opposed to applying for startup funding). A larger business will also not be tasking software developers with writing funding applications for external funding. Whether the lack of funding ultimately had notable bearing on the outcome of Startup A is impossible to determine. Past studies have suggested that venture capital has no significant effect on startup outcome [21] or may even affect it negatively [6].

Some of our findings in relation to the lack of resources experienced by software startups are novel. The struggles of Startup A to secure funding contributed to their lack of time resources. Forced to devote work hours towards applying for and otherwise seeking for funding, they found themselves with less time to work on furthering their business (PEC3). This provides one way in which the lack of resources experienced by software startups can be considered unique. Moreover, the lack of funding was also single most important source of uncertainty for Startup A (PEC4).

Finally, the inexperience of the founder and the team contributed to their perceived lack of resources (PEC1). As no one on the team had experience with entrepreneurship, they were forced to devote resources towards e.g. studying different legal company forms. An experienced startup founder will arguably find themselves devoting less time towards these types of activities, reducing the workload of the entire team, and lending more work hours towards productive activities. This supports extant literature which has linked business and technical experience with success in software startups [16, 24]. Our findings help us better understand *why* this is the case.

### 6.1 Limitations of the Study

This study was carried out as a single in-depth case study. This approach can be considered to present issues in terms of generizability of some claims, and further studies on the topic may offer different viewpoints. However, even a single case study can be enough to form a theory [13], and is especially appropriate for new topic areas [2], which was the case here.

We specifically wished to take on a more in-depth, exploratory approach to better understand the nature of uncertainty experienced by software startups. This study is a part of on-going research in which we apply the alternate templates sensemaking strategy described by Langley [13]. We plan to utilize other theories such as the Commitment Net Model of Abrahamsson [1] to analyze these data comparatively in future paper(s), using multiple background theories.

One further limitation in our data is also that the data were collected from the business-oriented founder of Startup A. As the data included few technical decisions related to the SE aspect of the startup, it is likely that some technical decisions made by the other team members were never discussed with the (non-technical) founder.

In terms of limitations, we would also like to draw attention to the Cynefin framework [12] used in the analysis of the data. The framework has some limitations that should be discussed: (1) the framework is a decision-making framework intended to help its users make sense of a situation in order to act accordingly rather than a categorizing tool for retrospective use as we have done here; (2) the subjective perception of an expert can make a complicated decision seem simple; and (3) the level of detail in the decisions greatly affects the level of orderedness associated with them.

---

[2] E.g., https://yle.fi/uutiset/3-10669492 ("More than 10 000 open programmer positions, but no one to fill them").

We have attempted to tackle the second limitation by having two researchers individually analyze the data, with a third one then weighing in on the conflicts. E.g. in some cases, the founder considered some complicated decisions to have been simple due to their in-depth knowledge of the business. Finally, in terms of the third limitation, in analyzing decisions using the Cynefin framework [12], one has to consider the level of analysis. A decision for the startup to acquire funding is a simple one: the startup needs capital to keep operating. However, the actual process of securing the funding can be anything from simple to chaotic.

### 6.2 Future Research Suggestions

This study aimed to validate one of the characteristics commonly attributed to software startups: uncertainty, which has even been described as chaotic. Various other characteristics have been attributed to software startups, of which a full list can be found in the introduction. Out of these characteristics, highly risky, lack of (SE) experience, innovativeness, and (external/market) time pressure appear to lack empirical support [9]. Whether startups really are unique in relation to these characteristics remains a question. E.g. most startups fail, but so do most new small businesses [9].

On a more general level, this on-going debate underlines the fact that software startups, and startups in general, lack a universally accepted definition. Thus, any research aiding the scientific community in defining software startups is welcomed to weigh in on this discussion.

## 7 Conclusions

In this paper, we have conducted a single, in-depth case study of a software startup. The case study was carried out in an ethnography-inspired fashion as one of the authors acted as the founder of a real-life startup. Data from the case were collected through video diary entries in which the founder described the decisions made by the startup. The aim of this case study was to study decision-making in software startups in relation to whether software startups operate amidst uncertainty, or even chaos.

We argue that software startups are in fact not characterized by a *unique* uncertainty that would notably differ from the uncertainty experienced by any other type of business organization. Software startups are largely faced with decisions that are analytical and can be solved by using existing good or best practices. Only a small portion (15,4%) of the decisions faced by the case startup required experimentation and could be considered to have involved notable levels of uncertainty. Much of the uncertainty in the decision-making in this case startup was related specifically to a lack of resources (and specifically funding). Mature firms also face issues with financial resources, although their ways of tackling them can differ (lay-offs etc.), leading us to question the uniqueness of software startups in relation to their alleged uncertainty.

To summarize our findings into practical implications, we underline the importance of: (i) understanding what capabilities are needed in the team, and aiming to secure them early on; and (ii) inexperienced software startup founders understanding the need to study various practical entrepreneurship skills, and doing so in an effective and structured manner via educational courses among other options.